\documentclass[pra,twocolumn,preprintnumbers,amsmath,amssymb,superscriptaddress,showpacs,longbibliography,normalem]{revtex4-2}

\RequirePackage{fix-cm}
\RequirePackage{amsmath}

\usepackage[T1]{fontenc} 
\usepackage[utf8]{inputenc} 

\usepackage{dsfont}
\usepackage{graphicx}
\usepackage{latexsym}
\usepackage{amsmath}
\usepackage{graphics}
\usepackage{amssymb}
\usepackage{layout}
\usepackage{verbatim}
\usepackage{amsfonts,epsfig}
\usepackage{soul}
\usepackage[dvipsnames,table]{xcolor} 
\usepackage{hyperref}
\usepackage{tikz}
\usepackage[compat=1.1.0]{tikz-feynman}
\usepackage{color}
\hypersetup{colorlinks, linkcolor={blue!90!black}, citecolor={blue!90!black}, urlcolor={blue!90!black} }
\usetikzlibrary{quantikz}

\usepackage{dsfont}

\newcommand{\beq}{\begin{equation}}
\newcommand{\eeq}{\end{equation}}
\newcommand{\bea}{\begin{eqnarray}}
\newcommand{\eea}{\end{eqnarray}}

\usepackage{color}

\usepackage{ulem} 

\begin{document}

\title{Signatures of quantum noise in the operation of Deutsch's algorithm}

\author{Ma{\l}gorzata Strza{\l}ka}
\affiliation{FZU - Institute of Physics of the Czech Academy of Sciences, Na Slovance 2, 182 00 Prague, Czech Republic}
\affiliation{Faculty  of  Mathematics  and  Physics, Charles University, Ke Karlovu 5,
121 16 Prague, Czech Republic}
\author{Katarzyna Roszak}
\affiliation{FZU - Institute of Physics of the Czech Academy of Sciences, Na Slovance 2,  182 00 Prague, Czech Republic}

\date{\today}

\begin{abstract}
We use Deutsch's algorithm as a stand in for more complex quantum algorithms in order to 
determine how quantum properties of an environment manifest themselves in results that can be
obtained on quantum computers. We model pure dephasing in two different ways; one keeps the 
full density matrix of the qubits and environments (quantum) 
while the other uses Kraus operators (classical).
We find that 
a single run of the algorithm yields the same effect in both cases, but running the algorithm twice leads to stark differences. Taking correlations and interplay between different decoherence
processes into account leads to a slowing of decoherence effects for balanced functions. For constant functions,
the effect is much more pronounced, and there is a qualitative change 
in the dependence of measurement outcomes on decoherence. We present results obtained on one of 
the IBM Quantum processors, which fully reproduce the predicted effect regardless of the 
assumptions made in the derivation. 
We further illustrate the findings on NV center spin qubits, which show more complex behavior
due to a small size of the environment.
\end{abstract}
\maketitle

\section{ Introduction}  
Quantum computers are a subject of enormous interest due to their exceptional potential to improve efficiency of certain types of computations, most notably search tasks \cite{grover98,jones98,giri17,wang20p}, and optimization problems \cite{ajagekar19,abbas24, wang23}.
Scalable quantum computers would allow for the
simulation of many physical systems enabling a more direct study of complex effects, potentially leading
to broad application in
chemistry and materials science. Notably, interesting studies of this type are currently
being performed on the quantum computer prototypes that are available, overcoming their 
limitations which are mostly due to a limited number of sparingly interconnected qubits
and noise \cite{bassman21,francis20}.

The development of quantum technology still faces many key challenges, such as scalability and fidelity of operation. The currently most advanced quantum computers use solid-state qubits, which means that 
interactions with some form of environment are unavoidable.
The viability of solid state qubits for utilitarian applications of quantum computing is thus intrinsically constrained by the duration over which they preserve coherent superpositions of quantum states \cite{Onizhuk25,abobeih22, mills22, rohling12}.
The predominant decoherence mechanism
for platforms based on solid-state qubits is pure dephasing (PD), meaning that 
the environment leads to diminished coherence, but cannot lead to relaxation, since 
energy exchange between the qubits and their environment is limited. 
This is the case for many types of solid state qubits in their regimes of operations,
including charge \cite{Liao10,salamon17,li22,ranni24}, spin \cite{yulin18,burkard23,khandelwal23,fung24,sieu25,beukers25}, and superconducting qubits \cite{bertet05,cywinski08,knee16,bravyi22,zhao22,anferov24}.

Due to the dominance of solid state qubits, the ubiquity of decoherence, and in light of the progress being made in the manufacturing 
and control of many-qubit systems, 
the study of the effects of decoherence becomes pressing. 
Modeling decoherence via reduced density matrix techniques, such as quantum channels \cite{nielsen00},
Kraus operators \cite{kraus83}, Lindblad master equations \cite{hall14}, but also more advanced methods from the theory
of open quantum systems \cite{nakajima58,zwanzig60,breuer02}, does not allow for the inclusion
of effects that stem from the build up of quantum correlations between the qubit system and 
the environment, or the transfer of information between the systems. Yet, it is known 
that these factors are relevant, as it has been shown that the fidelity of 
repeated teleportation is significantly affected by the build up of system-environment
correlations \cite{harlender22} and it is possible to reduce the impact of decoherence through re-teleportation for non-classical noise sources \cite{roszak23}.

We study the effect of decoherence on the operation of Deutsch's algorithm \cite{dj1992}
in order to detect and observe the signatures of quantumness in the noise, as well as to
determine when such effects are relevant. We apply methods which allow to track 
the full system-environment density matrix throughout the operation of the algorithm 
which are applicable as long as the interaction can only lead to pure dephasing of the qubits
\cite{roszak15,roszak18}. We compare these results to ones obtained when the same decoherence
is modeled via phase damping channels. Only the first type of modeling allows to take into 
account correlations formed between the system and the environment throughout the evolution,
as well as the effect that the qubits have on the state of the environment. Thus this method
allows for a more realistic description, including particularly quantum effects, such 
as destructive and constructive interference-like effects that can occur between
different decoherence processes \cite{roszak23}.

Firstly, we find that there is no difference between the two ways of modeling decoherence
if the algorithm is performed only once. This is in agreement with expectations, since similar
results were obtained in teleportation \cite{harlender22,roszak23}. The reasons are the same:
the nature of the dephasing mechanism cannot be determined by observation of the simple
decoherence of a qubit and in the case of the simplest quantum algorithms there is little difference between running the algorithm and just measuring coherence.

Yet if we run the algorithm twice, thus artificially complicating it in order for it to provide
a still simple demonstration of the operation of more sophisticated quantum programs, we observe
a stark difference between the two types of results. For balanced functions, the effect 
of quantum correlations building within the system and environment is limited to a slowing
of decoherence (which in itself may be useful, but is not very distinct). For constant functions
we observe a qualitative change in the behavior, which involves the change of the limiting
values of the probabilities of different measurement outcomes. This is a potentially 
important result, as it allows to witness the signatures of quantumness in noise
without any comparative data.  

The choice of Deutsch's algorithm, the simplified two-qubit version of the Deutsch-Jozsa 
algorithm \cite{dj1992},  for this study is due specifically to its simplicity. A double run
of the algorithm contains most of the elements which are required in ``serious'' algorithms, but it requires only two qubits. Two qubits with two environments are much more manageable 
and this facilitates the understanding of what underlies the observed changes in the 
effects of noise. 

We demonstrate our results on two different platforms. Firstly, we ran Deutsch's algorithm on
one of the IBM Quantum Heron r2 processors
(once to obtain decoherence factors for modeling with quantum channels and twice to obtain
quantum results). Here we only used two qubits and studied the behavior of the algorithm
under the effect of the natural decoherence that affects superconducting transmon qubits.
In order to vary the decoherence, the procedure was repeated for qubits that were increasingly
far apart. 
Quite surprisingly, we found results mirror theoretical predictions obtained assuming 
exponential pure dephasing, and show the characteristic behavior only present for constant
functions under quantum noise. This suggests that the decoherence present in the processor
is predominantly pure dephasing, but also that it is quantum in its nature and that the 
effect that the qubits have on the environment is non-negligible for further processing. 

The second example is a theoretical study of NV center spin qubits interacting with a 
nuclear environment \cite{doherty13,awschalom18,wood18,tchebotareva19,kuniej25}. It is chosen, because the decoherence is not exponential and the system is likely to demonstrate few-body effects. We find that decoherence curves can be substantially 
different in this case, but this is only highly visible for constant functions.
For balanced functions, the effect is limited to a slowing of decoherence, similarly as in 
all other cases that were studied. 

The paper is organized as follows. In Sec.~\ref{sec2} we outline Deutsch's algorithm. Sec.~\ref{sec3} describes the two ways of modeling pure dephasing. The effect of decoherence on Deutsch's algorithm
is discussed in Sec.~\ref{sec4} for one cycle of the algorithm, and in Sec.~\ref{sec5}
for two cycles. In Sec.~\ref{sec6} we present results obtained on the quantum computer.
In Sec.~\ref{sec7} we show theoretical results for NV center spin qubits in diamond. Sec.~\ref{sec8}
concludes the paper.

\section{Deutsch's algorithm\label{sec2}}

Deutsch's algorithm is the simplest boolean version of the Deutsch-Josa algorithm
\cite{dj1992}. The aim of the algorithm is to determine whether a function is constant or 
balanced, and in the version used here, both the input and the output of the function under
study, $y=f_n(x)$, are qubits (so $x=0,1$ and $y=0,1$). Classically, determining which type 
of function $f_n(x)$ is requires the function to be called twice, for $f_n(0)$ and $f_n(1)$,
yet Deutsch's algorithm provides a method which requires the function to be called only once.
There are obviously four possible functions $f_n(x)$ and the index $n=0,1,2,3$ serves to distinguish between them.

Running the algorithm requires only two qubits, and the key here is a two-qubit gate, $\hat{U}_{f_n}$,
that implements $f_n(x)$ as follows: $\hat{U}_{f_n}|a\rangle|b\rangle=|a\rangle|f(a)\oplus b\rangle$.
Here $a,b=0,1$ label the states of qubit $A$ and $B$, respectively.
These gates are explicitly given by
        \begin{subequations}
 \begin{eqnarray}
\hat{U}_{f_0}&=&\mathds{1},\label{f0}
\\\hat{U}_{f_1}&=&CNOT,\label{f1}
\\\hat{U}_{f_2}&=&(\hat{\sigma}_x\otimes \mathds{1}) CNOT (\hat{\sigma}_x\otimes \mathds{1}),\label{f2}
\\\hat{U}_{f_3}&=&\mathds{1}\otimes \hat{\sigma}_x,\label{f3}
\end{eqnarray}
\end{subequations}
where the labels $n=0,3$ correspond to constant functions, while $n=1,2$ correspond to 
balanced functions (and $CNOT$ denotes the controlled-NOT gate, while $\hat{\sigma}_x$ is the Pauli-$X$ gate). It is relevant to note here, that only for the balanced functions, the two-qubit gates are entangling gates,
as it will be relevant for the understanding of why the two types of functions 
are effected so differently by the noise. 
If both qubit input states for one of these gates are equal superposition states
$|\pm\rangle=(|0\rangle\pm |1\rangle)\sqrt{2}$
(with a plus and minus, respectively for qubit $A$ and $B$), then the information about 
the nature of function $f_n(x)$ is contained in the output state of qubit $A$:
it is $|+\rangle$ for constant functions, and $|-\rangle$ for balanced.

We do not include a flow chart of Deutsch's algorithm separately, but the whole algorithm with qubits 
starting in states $|0\rangle$ and $|1\rangle$ and measurement performend in the $\{|0\rangle,|1\rangle\}$ basis 
can be seen in Fig.~\ref{DA2}. If there is no decoherence (the environments $\hat{R}_A$ and $\hat{R}_B$ are disregarded),
then one cycle of the algorithm (everything before the vertical line) corresponds precisely to this flow chart.

\section{Modeling pure dephasing\label{sec3}}

We limit our study to decoherence of pure dephasing \cite{zurek03}, which
is ubiquitous for solid state qubits. The advantage of using this type of decoherence
from a computational perspective becomes obvious when modeling the operation of quantum
algorithms on the full system-environment density matrix. This is due to the existence of
a convenient formal solution for the evolution of this density matrix, which 
allows to keep track of the joint state when performing operations and measurements
solely on the qubits \cite{roszak15,roszak18}.

\subsection{Quantum environment\label{sec3a}}

To understand the full effect of a quantum environment on the operation of an algorithm,
it is often not enough to deal with the reduced density matrix of the qubit subsystem
\cite{harlender22,roszak23}. This is because decoherence is the result of correlations
being formed between the system and the environment \cite{roszak20}, both quantum and classical,
thus operations and measurements performed on the qubits, affect the environment, and
reciprocally, the changes in the environment are reflected in the state of the qubit.

When the interaction can only lead to pure dephasing of the qubits, then there must exist
a special basis in the qubit Hilbert space (called the pointer basis \cite{zurek03}), which 
diagonalizes the qubit Hamiltonian and the interaction. In this basis, denoted as $\{\ket{k}\}$,
the system-environment Hamiltonian can be written as \cite{roszak18}
\begin{equation}
\label{hamH}
\hat{H}_{\mathrm{PD}}=
\sum_{k}|k\rangle\langle k|\otimes\tilde{V}_k,
\end{equation}
where 
\begin{equation}
    \tilde{V}_k=\varepsilon_k\mathds{1}_{\mathrm{E}}+\hat{H}_{\mathrm{E}}+\hat{V}_k
\end{equation}
is the operator governing the evolution conditional on the pointer state
of the system. Here, $\varepsilon_k$ denote the eigenenergies of the free system Hamiltonian,
$\hat{H}_{\mathrm{E}}$ is the free Hamiltonian of the environment, and $\hat{V}_k$
are environmental operators in the interaction Hamiltonian (which has to be of the same
form as eq.~(\ref{hamH})).
Since this Hamiltonian is diagonal in the system subspace, the corresponding 
system-environment evolution 
operator can be written in a similar form
\begin{equation}
\label{UH}
\hat{U}(t)=
\sum_{k}|k\rangle\langle k|\otimes\hat{w}_k(t),
\end{equation}
where $
    \hat{w}_k(t) = e^{-\frac{i}{\hbar}\tilde{V}_k t}$.

In the following, we assume that each qubit, $A$ and $B$, has a separate environment,
so we are not taking into account any interaction between the qubits facilitated by
the environment \cite{krzywda16}, even though the method allows for this to be done. 
Thus the 
pure dephasing Hamiltonian is of the form $\hat{H}_{\mathrm{PD}}=\hat{H}_{A}+\hat{H}_{B}$,
where both $\hat{H}_{A}$ and $\hat{H}_{B}$ have the form of eq.~(\ref{hamH}), so 
the evolution operator for the two qubits and the environment is given by
  \begin{equation}
      \label{u2}
      \hat{U}(t)=\sum_{i,j=0,1}\ket{ij}\bra{ij}\otimes\hat{w}_i^A(t)\otimes\hat{w}_j^B(t).
  \end{equation}
Here $i$ and $j$ label the pointer states of qubit $A$ and qubit $B$, respectively, and 
the conditional evolution operators acting on the environments corresponding 
to each qubit are denoted by $\hat{w}_{i/j}^{A/B}(t)$. The unitary operators
\begin{equation}
\label{wj}
\hat{w}_{i/j}^{A/B}(t)=e^{-\frac{i}{\hbar}\varepsilon_{i/j}^{A/B} t}e^{-\frac{i}{\hbar}(\hat{H}_{\mathrm{E}}^{A/B}+\hat{V}_{i/j}^{A/B})t}
\end{equation}
contain information not only about the evolution of the environment, but also about
the free evolution of the qubits. We will not be taking the free qubit evolution into account,
since we are only interested in the effects of the environment on algorithm operation.

In modeling decoherence, we also assume that there are no initial correlations present between the environments (or between
the environments and the qubits), so they are initially in a product state,  $\hat{R}_A(0)\otimes\hat{R}_B(0)$.

\subsection{Classical decoherence\label{sec3b}}

Typically, in the study of the effects of noise on algorithm operation, decoherence channels are
used \cite{nielsen00}. These do not allow to capture the effects stemming from the build up of correlations, 
or the effect that the qubit has on the state of the environment,
but they provide a straightforward and elegant method for the description of the
reduced density matrix of the qubit system. We focus on the phase damping channel, as
it describes the same type of process as is described by PD Hamiltonians.

In the operator-sum representation, the evolution of the state of two qubits under 
the effects of local noise can be written as
\begin{equation}
      \epsilon(\hat{\rho})=\sum_{i,j=0,1}(E_i^A \otimes E_j^B)\hat{\rho} (E_i^A \otimes E_j^B),
\end{equation}
with Kraus operators
\begin{subequations}
    \begin{eqnarray}
        E_0^{A/B}&=&\left[\begin{array}{cc}
            1 & 0 \\
            0 & c_{A/B}
        \end{array}\right],\\
     E_1^{A/B}&=&\left[\begin{array}{cc}
            0 & 0 \\
            0 & \sqrt{1-c_{A/B}^2}
        \end{array}\right],
    \end{eqnarray}
\end{subequations}
which satisfy the trace-preserving condition $ \sum_{i,j}(E_i^A\otimes E_j^B)^*(E_i^A\otimes E_j^B)=\mathds{1}$.

In many situations, Kraus operators can be used to describe noise stemming from interactions with a quantum
environment. Under the assumption that the correlations formed between the system and the 
environment undergo rapid decay (so they are of no relevance in the second decoherence process)
and that the effect of the interaction on the system state is negligible,
the decoherence factors $c_{A/B}$ can be obtained from the evolution operators (\ref{u2}),
\begin{equation}
\label{ca}
    c_{A/B}= \langle \left(\hat{w}_{1}^{A/B}(t)\right)^{\dagger}\hat{w}_{0}^{A/B}(t)\rangle,
\end{equation}
where the expectation value is taken with respect to the initial state of the appropriate 
environment ($A$ or $B$).

\begin{widetext}
\begin{center}
    \begin{figure}[!t]
	\centering
   \includegraphics[width=1.\linewidth]{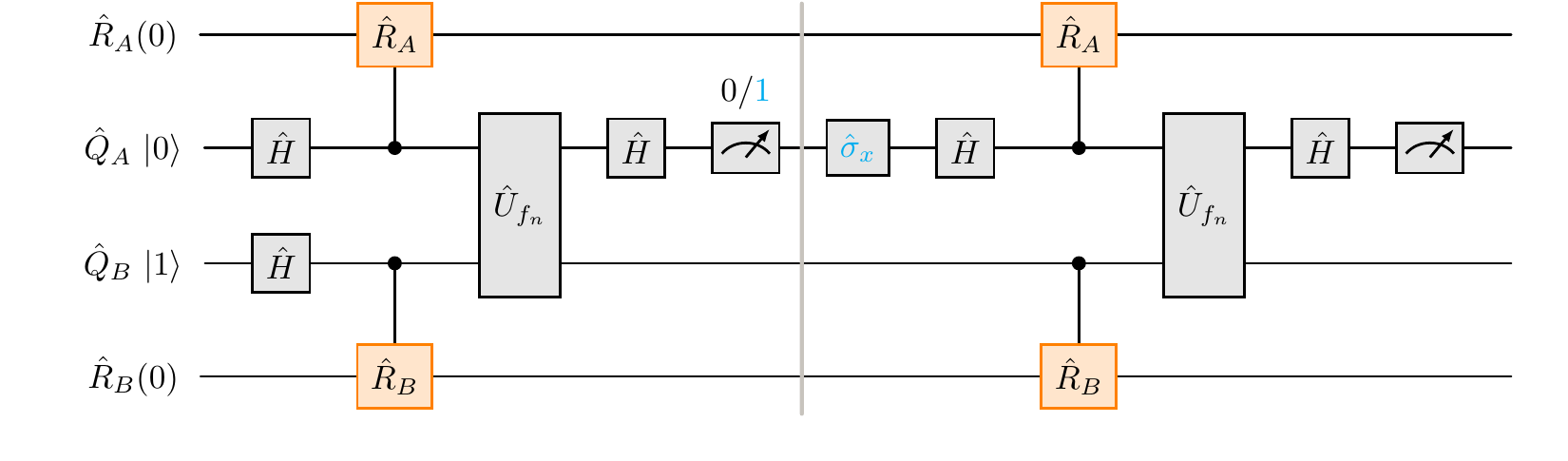}
	\caption{Two cycles of Deutsch's algorithm with decoherence. The end of the first run of the algorithm
    ends with the measurement of qubit $A$ and is denoted by the vertical line. The Pauli-X gate is applied on qubit
    $A$ only if 
    the measurement outcome is $|1\rangle$.
    Both qubits undergo decoherence
    right before the application of the two-qubit unitary functions $\hat{U}_{f_n}$. The decoherence process
    with the same environments is repeated in both cycles. 
		\label{DA2}}
\end{figure}
\end{center}
\end{widetext}

\section{Deutsch's algorithm under pure dephasing\label{sec4}}

In the following, we consider the operation of Deutsch's algorithm under noisy conditions 
in such a way, that the decoherence affects the qubits directly before the operation of the two-qubit gate that implements the (constant or balanced) function $f_n(x)$. 
Fig.~\ref{DA2} depicts two cycles of Deutsch's algorithm, both of which undergo decoherence. The end of the first 
cycle, which involves a measurement on qubit $A$ is denoted by the vertical line.

If the algorithm is ran, as intended, only once then the nature of the decoherence plays no role
on its operation.
For constant functions, realized by unitary transformations given by Eqs (\ref{f0}) and (\ref{f3})), the probability of measuring $|0\rangle$ on qubit $A$ at the end (which is the correct measurement result) is given by
\begin{equation}
    p_0^{f_{0/3}}=\frac{1+c_A}{2}.
    \label{ch_p0c}\end{equation}
For balanced functions (realized by eqs (\ref{f1}) and (\ref{f2})), the probability of measuring $|0\rangle$
(the wrong result) is
\begin{equation}
    p_0^{f_{1/2}}=\frac{1-c_Ac_B}{2}.
    \label{ch_p0b}
\end{equation}
Obviously, the probabilities of measuring $|1\rangle$ for all functions is  $p_1^{f_{n}}=1-p_0^{f_{n}}$.

Relevantly, but unsurprisingly, only decoherence of qubt $A$ affects the operation of the algorithm for 
constant functions, because in this situation all operations are performed separately on both qubits. 
For balanced functions, decoherence on both qubits is relevant, since the two-qubit gates that implement
the functions $f_{1/2}(x)$ are actual controlled two-qubit gates.  

\section{Two cycles of Deutsch's algorithm under decoherence\label{sec5}}
Similarly as in the case of teleportation \cite{harlender22,roszak23}, the protocol needs to be ran twice for
the nature of the environment to play a role in how it effects algorithm operation. 
In the following, the function $f_n(x)$ will not be changed between the cycles, since we are studying how
the effect 
of decoherence can be diminished by the repetition. Furthermore, for simplicity we will assume that the level of decoherence
introduced in both steps is the same.

The two cycles are depicted in Fig.~\ref{DA2}, where the realizations of Deutsch's algorithm are separated
by a vertical line. The second cycle begins with a conditonal operation on qubit $A$, so that the initial
state of the qubit is $|0\rangle$ regardless of the measurement outcome (a Pauli-X gate is applied if the 
measurement outcome is $|1\rangle$). Furthermore, in the second cycle there is no Hadamard gate on qubit $B$,
because the qubit is already in the $|-\rangle$ state after the first cycle assuming ideal opeartion (no decoherence).

In the following we find the probabilities of the outcome of the second measurement on qubit $A$
conditional on the outcome of the first measurement, denoted as $p_{\alpha,ij}^{f_n}$ (where $i=0,1$ is the outcome
of the first measurement and $j=0,1$ corresponds to the second measurement, and $\alpha$ distinguishes the type of 
decoherence process). Thus to find the probabilities of obtaining measurement outcomes $i$ and $j$ in the two measurements, it is necessary to multiply the conditional
probability by the corresponding probability obtained in a single cycle of the algorithm 
(given by eqs (\ref{ch_p0c}) and (\ref{ch_p0b})), $P_{\alpha,ij}^{f_n}=p_{i}^{f_n}p_{\alpha,ij}^{f_n}$.

\subsection{Classical decoherence\label{sec5a}}

Using Kraus operators to model decoherence makes introducing dephasing in the second cycle particularly simple.
After two cycles of the algorithm, for constant functions we obtain the probability of a correct measurement
outcome conditional on the correct measurement outcome in the first step given by
\begin{equation}
    p_{c,00}^{f_{0/3}}=\frac{1}{2}+\frac{c_A}{2},
\label{ch_p00cc}\end{equation}
and the probability of an incorrect outcome given an incorrect outcome was obtained in the first step,
\begin{equation}
    p_{c,11}^{f_{0/3}}=\frac{1}{2}-\frac{c_A}{2}.
\label{ch_p11cc}\end{equation}
The other two probabilities can be easily obtained, since 
$p_{\alpha,01}^{f_{n}}=1-p_{\alpha,00}^{f_{n}}$
and $p_{\alpha,10}^{f_{n}}=1-p_{\alpha,11}^{f_{n}}$.
For these functions, the conditional probabilities with the correct/incorrect outcome in the second measurement
are the same as the probabilities of the correct/incorrect outcomes of the first measurement, eq.~(\ref{ch_p0c}).
This is to be expected, since the first measurement forces the return of qubit $A$ to a pure state,
and the evolution of qubit $B$ has no bearing on the algorithm for constant functions. 

For balanced functions, the corresponding conditional probabilities are more complicated, since qubit $B$
is in a mixed state at the beginning of the second step. They are given by
\begin{eqnarray}
    \label{ch_p00bb}
    p_{c,00}^{f_{1/2}}&=&\frac{1}{2}+\frac{c_Ac_B(c_A-c_B)}{2(1-c_Ac_B)},\\
    p_{c,11}^{f_{1/2}}&=&\frac{1}{2}+\frac{c_Ac_B(c_A+c_B)}{2(1+c_Ac_B)}.
\label{ch_p11bb}
\end{eqnarray}

\subsection{Quantum decoherence\label{sec5b}}
The situation is more complicated when modeling decoherence with the use of a quantum environment.
This is because keeping track of the build-up of correlations between the qubits and their respective environments
requires the full density matrix and the trace over the degrees of freedom of the environments can 
only be performed after the second decoherence process. In the case of pure decoherence this process is manageable
because of the simple structure of the formal solution for the evolution of the system-environment density matrix,
as outlined in Sec.~\ref{sec3a}.

Decoherence factors that come from a single dephasing process have already been defined in Sec.~\ref{sec3b}
and are given by eq.~(\ref{ca}). However, after two cycles of Deutsch's algorithm, with decoherence processes
that are the result of a qubit interacting with the same environment twice, these factors are no longer 
sufficient to describe observed decoherence. 
Analogously to what was observed for repeated teleportation \cite{harlender22,roszak23},
there are situations when the conditional evolution operators of the environment, eq.~(\ref{wj}), cancel out,
as well as situations when the decoherence processes are not independent and cannot be described as products
of single-run decoherence factors. Thus, for a full description additional decoherence factors are needed and they
are given by
\begin{equation}
    \label{ca2}
    d_{A/B}^2= \mathrm{Re}\langle \left(\hat{w}_{1}^{A/B}(t)\right)^{\dagger}\left(\hat{w}_{1}'^{A/B}(t)\right)^{\dagger}\hat{w}_{0}'^{A/B}(t)\hat{w}_{0}^{A/B}(t)\rangle,
\end{equation}
where the prime denotes the operators pertaining to the second decoherence process.
Under the assumption that the two processes are the same, the formula simplifies
\begin{equation}
    \label{c2tylda}
    d_{A/B}^2= \mathrm{Re}\langle \left(\hat{w}_{1}^{A/B}(2t)\right)^{\dagger}\hat{w}_{0}^{A/B}(2t)\rangle,
\end{equation}
and the factor is dependent on the decoherence that would be obtained by a single dephasing process 
that took twice as long. 

For constant functions, the conditional probabilities analogous to eqs (\ref{ch_p00cc}) and (\ref{ch_p11cc}) are given by
\begin{eqnarray}
\label{q_p00cc}
p_{q,00}^{f_{0/3}}&=&\frac{3+4c_A+d_A^2}{4(1+c_A)},\\
\label{q_p11cc}
p_{q,11}^{f_{0/3}}&=&\frac{3-4c_A+d_A^2}{4(1-c_A)}.
\end{eqnarray}
Although they are significantly more complicated, the probabilities still depend only on the decoherence 
of qubit $A$. Yet, there is a qualitative difference between the classical and quantum cases.
Classically, complete decoherence ($c_A=0$ and $d^2_A=0$) yields all conditional probabilities equal to $1/2$, while quantumly the probabilities of obtaining a repeat result are increased to $3/4$, while the probabilities of getting a different result are diminished to $1/4$. 

For balanced functions, the conditional probabilities corresponding to eqs  (\ref{ch_p00bb}) and (\ref{ch_p11bb}) are 
\begin{eqnarray}
\label{q_p00bb}
p_{q,00}^{f_{1/2}}&=&\frac{2-2c_Ac_B-c_A+c_B-c_Ad_B^2+d_A^2 c_B}{4(1-c_A c_B)},\\
\label{q_p11}
p_{q,11}^{f_{1/2}}&=&\frac{2+2c_Ac_B+c_A+c_B+c_Ad_B^2+d_A^2 c_B}{4(1+c_A c_B)}.
\end{eqnarray}
The functions are more complicated, but they depend on decoherence
of both qubits as in the classical case. Yet, contrarily to constant functions, there is no difference when 
the qubits are completely dephased, and all conditional probabilities are equal to $1/2$ for $c_{A/B}=0$ and $d^2_{A/B}=0$,
as in the classical case.

In order to make a quantitative comparison of the effects that quantum and classical decoherence
has on the operation of the algorithm for the two types of functions, some assumptions on the type of dephasing
process need to be made. If we assume that $d_{A/B}^2 = c_{A/B}^2$, which corresponds to exponential decay
according to eq.~(\ref{c2tylda}), then we obtain much simplified expressions for eqs (\ref{q_p00cc})-(\ref{q_p11}),
\begin{eqnarray}
\label{quant_apro1}
   p_{a,00}^{f_{0/3}}&=&\frac{1}{2}+\frac{1+c_A}{4},\\
\label{quant_apro2}
  p_{a,11}^{f_{0/3}}&=&\frac{1}{2}+\frac{1-c_A}{4},\\
\label{quant_apro3}
     p_{a,00}^{f_{1/2}}
     &=&\frac{1}{2}-\frac{c_A-c_B}{4},
        \\
\label{quant_apro4}
 p_{a,11}^{f_{1/2}}&=&\frac{1}{2}+\frac{c_A+c_B}{4}.
    \end{eqnarray}

\begin{figure}[!t]
    \centering
    \includegraphics[width=1.\linewidth]{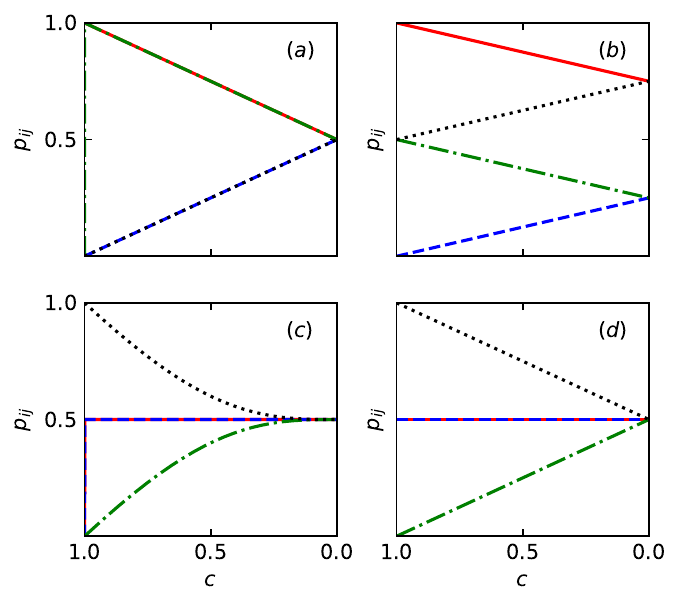}
    \caption{Conditional probabilities of measurement outcome after two cycles of Deutsch's algorithm as a function of coherence parameter $c$ ($p_{00}$ - solid red lines; $p_{01}$- dashed blue lines;  $p_{10}$ - dotted-dashed green lines; $p_{11}$ - dottedu black lines). Left panels correspond to phase damping and right panels to quantum decoherence. Top panels show results for 
    constant functions and bottom for balanced.}
    \label{cAB}
\end{figure} 

The conditional probabilities of the second measurement outcome as a function of the coherence parameter $c$ are shown in Fig.~\ref{cAB}, under the assumption that both decoherence processes are the same, $c_A=c_B=c$. 
Top panels show results for a constant function with classical decoherence (left), and quantum decoherence (right).
Bottom panels show analogous results for balanced functions. 

It is evident that the quantumness of the decoherence process
has a starkly different effect on the curves for constant and balanced functions. For balanced functions (bottom panels), the
effect on the probability of obtaining a correct measurement outcome, $|1\rangle$, after a correct result in the first run
(dotted black line) is slower. Because of the assumption that the decoherence is the same for both qubits, there is no $c$-dependence
of the conditional probabilities after an incorrect measurement outcome in the first run, and this does not change for quantum
decoherence. For constant functions (top panels), the quantum nature of the decoherence affects all conditional probabilities,
and it changes their dependence qualitatively. In the classical case, all probabilities are equal to $1/2$ for complete dephasing,
$c=0$, while in the quantum case, the conditional probabilities of obtaining the same result in the second measurement 
as was obtained in the first measurement are equal to $3/4$ for $c=0$, while obtaining a different result has a much lesser probability
of $1/4$.

For completeness and because they are more experimentally accessible, in Fig.~\ref{cAB_pipij} we plot the probabilities, not conditional, for obtaining given sets of measurement outcomes. Also here it is evident that for balanced functions, quantum 
effects only yield a reduced effect of the decoherence, while for constant functions there is a qualitative change. Only for 
constant functions are the probabilities of obtaining the same outcome heightened, while the probabilities of different outcomes
are lowered
due to the interference between the two decoherence processes.

\section{Quantum computer results \label{sec6}}

Here we present results obtained on the processor
ibm\textunderscore marrakesh, which is one of the IBM Quantum Heron r2 processors.
Existing quantum computer prototypes are known to be noisy \cite{murali19,preskill18, wang20p} and we performed a single and double run of Deutsch's algorithm
in order to study the nature of this noise. The single run of the algorithm served to determine the decoherence factors 
$c_A$ and $c_B$ from the measured probabilities using eqs (\ref{ch_p0c}) and (\ref{ch_p0b}). These factors are then used
to find the predicted conditional probabilities for classical decoherence from eqs (\ref{ch_p00cc})-(\ref{ch_p11bb}).
Quantum results are obtained directly by running the algorithm twice.

\begin{figure}[!t]
    \centering
    \includegraphics[width=1.\linewidth]{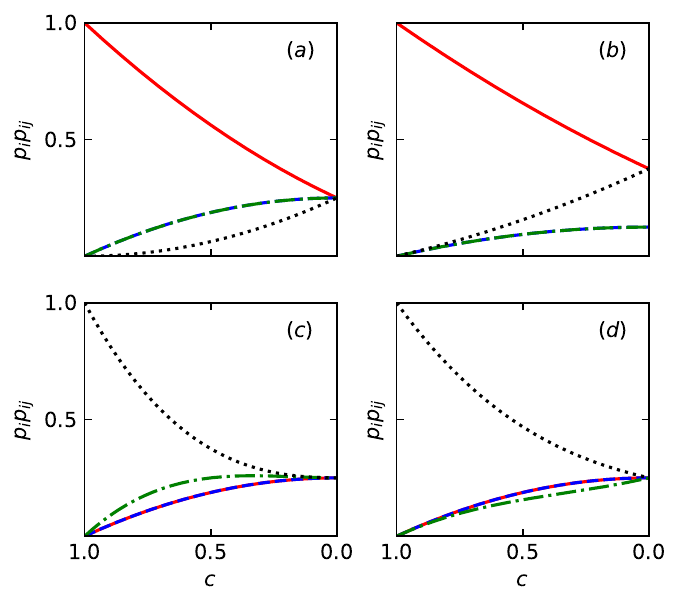}
    \caption{Probabilities of measurement outcome after two cycles of Deutsch's algorithm as a function of coherence parameter $c$ ($p_{00}$ - solid red lines; $p_{01}$- dashed blue lines;  $p_{10}$ - dotted-dashed green lines; $p_{11}$ - dotted, black lines). Left panels correspond to phase damping and right panels to quantum decoherence. Top panels show results for 
    constant functions and bottom for balanced.}
    \label{cAB_pipij}
\end{figure}

\begin{figure}[!t]
    \centering
    \includegraphics[width=1.\linewidth]{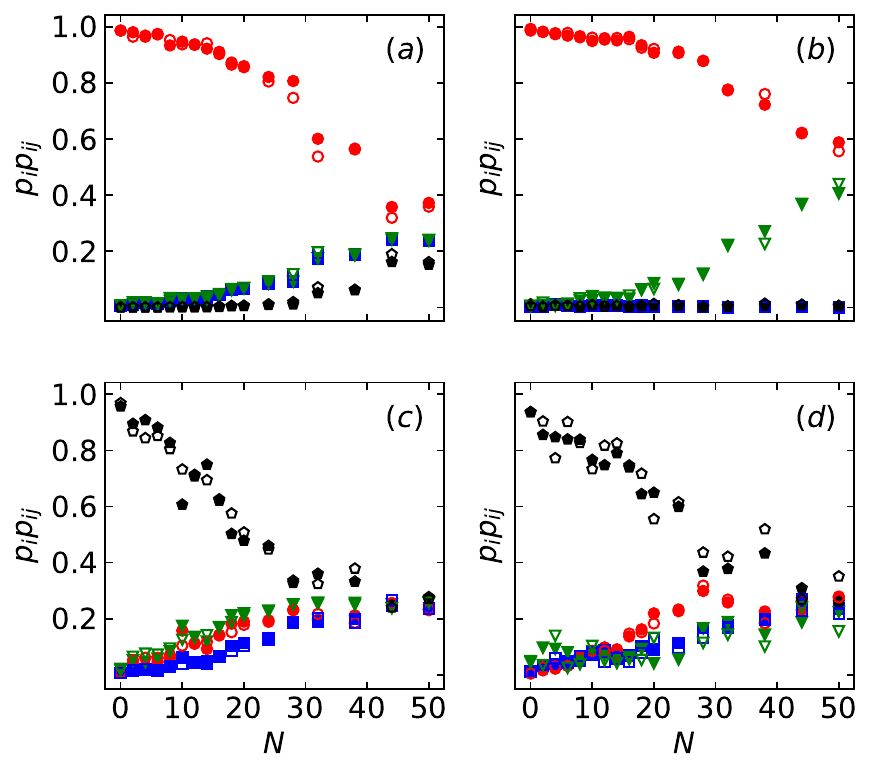}
    \caption{Probabilities of measurement outcome after two cycles of Deutsch's algorithm performed on the quantum processor
    as a function of the number of qubits between operational qubits ($p_{00}$ - red circles; $p_{01}$- blue squared;  $p_{10}$ - green triangles; $p_{11}$ - black pentagons; filled markers always denote functions $f_{0/1}$, while not filled denote
    $f_{2/3}$). 
    Panels on the left show probabilities obtained by decoherence channels, and panels on the right obtained directly on quantum computer. As previously, top panels show results for 
    constant functions and bottom for balanced. }
    \label{q_comp}
\end{figure}

In order to vary the degree of decoherence we run the algorithm using non-neighboring qubits
(details can be found in the Appendix, including an illustration of the circuit used). Predictably, the amount of decoherence grows with the number of qubits between them.
The results obtained on the quantum computer are plotted in Fig.~\ref{q_comp}, which is analogous to the neighboring Fig.~\ref{cAB_pipij}. The probabilities are plotted as a function of the number of qubits between the qubits on which Deutsch's
algorithm is operated, $N$. There are slight deviations in the results for the different types of constant and balanced functions,
so filled markers always correspond to functions $f_0$ (constant) and $f_1$ (balanced), while $f_2$ (balanced) and $f_3$
(constant) have not filled markers. 
Each data point is obtained by repeating the experiment a $1000$ times.  

The observed behavior 
is qualitatively very similar to the theoretically obtained results: For balanced functions (bottom panels), the effect of decoherence
is slower for a double run of the algorithm in comparison to the results obtained using the formulas for classical decoherence
in a double run
by extracting the decoherence factors from a single run of the algorithm. The more distinct predicted features for constant functions
are also present (top panels). The probabilities measured when running the algorithm twice show the division between
probabilities with the same outcomes and probabilities with different outcomes; the former converge to a higher level than 
obtained by the classical formulas, while the latter have distinctly smaller limiting values. This signifies that the 
decoherence mechanisms that affect qubits in the particular realization of the quantum computer that was used
are quantum to a high extent.

The strong qualitative agreement between the results obtained on the quantum computer and the theoretical prediction 
are surprising, because of the number of assumptions that were made in the modeling. The most important are the limitation
of the decoherence processes to pure dephasing and the assumption that the decoherence takes place only directly before 
the two-qubit unitary gate. Furthermore, the results suggest that the approximation $d_{A/B}^2 = c_{A/B}^2$
holds rather well. 
This last approximation can be strongly violated in other physical systems as we will demonstrate in the next section, thus the fact that no oscillations
are seen in the data suggests that the environment of the qubits is reasonably large. 

\section{NV center spin qubits interacting with nuclear environment \label{sec7}}

In order to demonstrate the quantum effects which can arise for pure dephasing that is not
exponential, meaning that $d^2_{A/B}\neq c^2_{A/B}$ and the approximate eqs (\ref{quant_apro1})-(\ref{quant_apro4}) cannot be used, here we provide the results for an NV spin qubit interacting
with a nuclear environment \cite{childress06,zhao12,kwiatkowski18,kwiatkowski20}. 
The NV center spin qubit is a defect in the diamond crystal lattice in which two neighboring carbon atoms are replaced by a nitrogen atom and a vacancy (a missing carbon atom). The lowest energy level of the NV center is formed by the spin qutrit, $S=1$.
The system under study consists of two NV center spin qubits, defined on the $S=0$ and $S=1$
spin states. Each qubit interacts with it's own environment of partially polarized 
nuclei of the $^{13}$C carbon isotope \cite{london13,fischer13,pagliero18,wunderlich17,scheuer17,poggiali17,hovav18}.
This environment is sparse, thus providing a perfect testing ground for exotic behaviors
in decoherence. 

Since we are only interested in the effects of decoherence, in the following we omit the free evolution of the qubit. For easy comparison with results obtained via the simplified formulas, 
we also assume that the environments are identical (with the same initial states and interactions).
For each qubit and environment, we only take into account the free
Hamiltonian of the environment,
	\begin{equation}
	\label{HE}
    \hat{H}_E=\sum_{k}\gamma_{n}B_{z}\hat{I}^{z}_{k},
	\end{equation}
    and the interaction term,
\begin{equation}
    \hat{H}_I=\ket{1}\bra{1} \otimes\sum_{k}\left( \mathbb{A}^{z,x}_k\hat{I}^{x}_{k}
	+\mathbb{A}^{z,y}_k\hat{I}^{y}_{k}+\mathbb{A}^{z,z}_k\hat{I}^{z}_{k}
	\right),
\end{equation}
where the coupling constants for each direction are given by
\begin{equation}
	\label{a}
	\mathbb{A}^{z,i}_k=\frac{\mu_0}{4\pi}\frac{\gamma_e\gamma_n}{r_{k}^3}
	\left(1-\frac{3(\mathbf{r}_{k}\cdot\hat{\mathbf{i}})(\mathbf{r}_{k}\cdot\hat{\mathbf{z}})}{r_{k}^2}
	\right).
	\end{equation}
    Here,
the nuclear spins are labeled by the index $k$ and $\hat{\mathbf{I}}_{k}$ is the spin 
operator for spin $k$.
$B_{z}$ denotes the magnetic field which is applied in the $z$ direction. 
The electron gyromagnetic ratio is $\gamma_e=28.08$ MHz/T and
the gyromagnetic ratio for spinful carbon $^{13}$C is $\gamma_{n} \! = \! 10.71$ MHz/T.
The displacement vector between the $k$-th nucleus and the qubit
is labeled by $\mathbf{r}_{k}$, 
$\hat{\mathbf{i}}=\hat{\mathbf{x}},\hat{\mathbf{y}},\hat{\mathbf{z}}$
	are unit vectors in the three distinct directions, and $\mu_0$ is the magnetic permeability of the vacuum.
The conditional environmental operators (\ref{wj}) that correspond to this Hamiltonian are explicitly given in Ref.~\cite{strzalka21}.

We assume that the initial state of each environment 
is partially polarized and uncorrelated and is given by $\hat{R}(0)  =  \bigotimes_{k} \hat{\rho}_{k}(0)$, where $\hat{\rho}_{k}(0)$ is the density matrix of nuclear spin $k$, 
	\begin{equation}
	\label{ini}
	 \hat{\rho}_k(0) =\frac{1}{2}(\mathds{1} + p_{k}\hat{I}^{z}_{k}).
	\end{equation} 
	Here $p_k \! \in \! [-1,1]$ is the polarization of the $k$-th spin.

\begin{figure}[!t]
    \centering
\includegraphics[width=1.\linewidth]{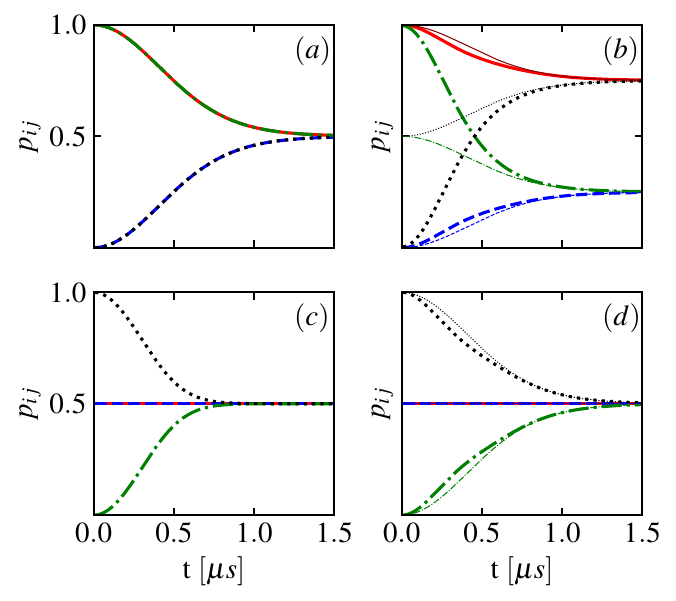}
    \caption{Conditional probabilities of measurement outcome after two cycles of Deutsch's algorithm on NV center spin qubits as a function of decoherence time. 
    Initial polarization of nuclear environment is $p=0.1$ and applied magnetic field is $B_z=0.1$ T.
    Individual probabilities are given by: $p_{00}$ - solid red lines; $p_{01}$- dashed blue lines;  $p_{10}$ - dotted-dashed green lines; $p_{11}$ - dotted black lines. Left panels correspond to phase damping and right panels to quantum decoherence. Top panels show results for 
    constant functions and bottom for balanced.
    Panels $(b)$ and $(d)$ additionally show results for $d_{A/B}^2 = c_{A/B}^2$ marked with thin 
    {\color{gray}grey} lines.}
    \label{p01b01}
\end{figure}

\begin{figure}[!t]
    \centering
\includegraphics[width=1.\linewidth]{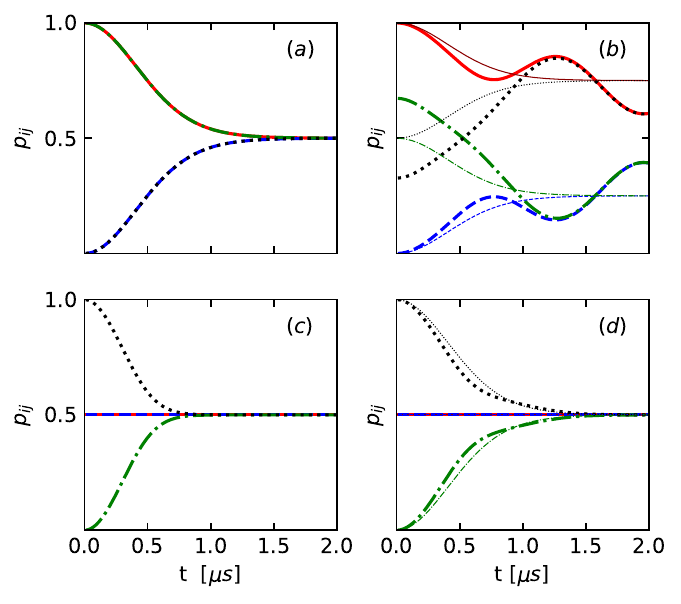}
    \caption{As Fig.~\ref{p01b01}, but for initially fully polarized nuclear spins, $p=1$.}
    \label{p1b01}
\end{figure}

Figs \ref{p01b01} and \ref{p1b01} show conditional probabilities of measurement outcome after two cycles of Deutsch's algorithm for
NV center spin qubits each interacting with 32 nuclear spins
(the table of coupling constants that were used can be found in the appendix)
as a function of decoherence time (which is assumed equal for both processes). 
These figures are analogous to Fig.~\ref{cAB}, but the quantum results were obtained using the full equations 
(\ref{q_p00cc})-(\ref{q_p11}). Curves that are obtained via the simplified equations
(\ref{quant_apro1})-(\ref{quant_apro4}) are plotted in the right panels with thin lines,
for comparison. The two figures differ only by the initial state of the environments;
in Fig.~\ref{p01b01} each spin is only partially polarized with $p_k=p=0.1$, while 
in Fig.~\ref{p1b01} all spins are fully polarized, $p_k=p=1$.
The applied magnetic field is equal to $B_z=0.1$ T in both figures.

As previously, for balanced functions (bottom panels) the effect of the fully quantum description of the noise (right) is limited to a diminished effect of decoherence, which takes twice as long 
to erase any information about the nature of the function under study. Here the difference
between curves obtained by the full and simplified equations is minimal. 

Contrarily, for 
constant functions (top panels) there is a stark difference between evolutions obtained
by the different sets of equations. The splitting 
of the value of different conditional probabilities for fully decohered states,
which is the most pronounced feature in Fig.~\ref{cAB} (b), is still present in the long time
behavior in both cases. It is the short time behavior (high value of coherence) that is qualitatively 
different. In fact, short-time behavior of the $p_{10}$ and $p_{11}$ curves 
(the probabilities of the correct and incorrect outcome, respectively, conditional on 
the first measurement having the incorrect outcome) resembles classical behavior. 
This is much more pronounced for low initial polarization of the environment (Fig.~\ref{p01b01}), since
an unpolarized environment is bound to behave more classically than a more polarized one
(since full polarization means that the environment is initially in a pure state).
A similar effect is responsible for the oscillations in the long time behavior
of all conditional probabilities visible in Fig.~\ref{p1b01} (b), which are the effect of the 
interplay between the two decoherence processes that occur on the qubits.

For completeness, we provide plots for non-conditional probabilities analogous to Fig.~\ref{cAB_pipij} and corresponding to Figs \ref{p01b01} and \ref{p1b01} in the Appendix. 
There it can be seen that all of the special features visible in the dependence of the 
conditional probabilities are also clearly visible on full probabilities.

\section{Conclusion \label{sec8}}
Decoherence is a common effect in quantum systems, especially in the solid state.
As such, it needs to be taken into account when modeling the running of algorithms
on quantum processors. Since algorithms are complex, often many stage schemes requiring
repeated application of gates and measurements, it is critical to be able to take into 
account the effects of quantumness of noise sources. 

We used Deutsch's algorithm as a prototypical example of a quantum algorithm. We modeled 
pure dephasing in a way that used the full density matrix of qubits and their environments throughout the run of the algorithm and compared the effect of decoherence with 
results for the same decoherence included using Kraus operators. We found that the nature 
of the noise starts playing a role only when the algorithm is ran at least twice. 
For two cycles of Deutsch's algorithm, the effects of qubit-environment correlations 
and environmental memory are distinct. For balanced functions, we observe a slowing of 
the effect of decoherence with respect to the classical counterpart. Yet for constant
functions, there is a qualitative change between the two types of modeling. Quantum 
environments lead in this case to a change in the value of probabilities 
of given measurement outcomes, so the nature of the noise can be determined without 
comparison to classical expectations. 

We tested the results using an IBM Quantum processor. Here the simple two-qubit algorithm
was ran twice, and the noise was supplied by the natural environment present for superconducting
qubits that form the computer. The amount of decoherence was varied by changing the distance
between the qubits that were used in Deutsch's algorithm. 
We observed distinctly similar results to the theoretical predictions, including 
the behavior characteristic only for constant functions under full noise modeling. 
This suggests that the environments for the specific quantum processor qubits retain
information about the qubit state (obtained due to qubit-environment interaction)
longer than gate operation times. Taking these features of the noise into account 
is relevant for proper noise mitigation when running experiments on quantum
computers.

We also presented results for NV center spin qubits in a diamond lattice 
interacting with an environment of $^{13}$C spinful nuclei. Here the environments are sparse,
so additional effects due to non-exponential decoherence were observed, demonstrating
the variety of possible effects of decoherence on algorithm operation. 
As previously, the effects are most distinct for constant functions.

\section*{Acknowledgment}
MS: This work was co-funded by the European Union and
the Czech Ministry of Education, Youth and Sports
(Project TERAFIT – CZ.02.01.01/00/22\_008/0004594) and the Grant Agency of Charles University (122124).
K.R. and M.S: This project is funded within the QuantERA II Programme that has received funding from the EU H2020 research and innovation programme under GA No 101017733, and with funding organization MEYS Czech Republic.  
Computational resources were provided by the e-INFRA CZ project (ID:90254),
supported by the Ministry of Education, Youth and Sports of the Czech Republic.
This research was supported by the Quantum Innovation Center (QIC) project under the Consortium Agreement for the Use of Quantum Technology Services No. UO/09/007/2025. We acknowledge the use of IBM Quantum services for this work. The views expressed are those of the authors, and do not reflect the official policy or position of IBM or the IBM Quantum team. In this paper we used 
ibm\textunderscore marrakesh, which is one of the IBM Quantum Heron r2 processors.


%

\appendix
\section{Details for quantum computer results}

The experiment on the quantum computer starts with run of single and double cycles of Deutsch's algorithm on two neighboring qubits.
On a quantum computer, the qubits are initially in the $\ket{0}$ state, therefore, the  Pauli-X gate is used in order to initialize the $\ket{1}$ state on qubit $B$, after which the algorithm proceeds as described in the main body of the article.
In order to vary the level of decoherence, the number of qubits beetween the operational qubits on which the algorithm was run
is changed. This is achieved by the application of swap gates
on qubits $A$ and $B$ and their neighboring qubits $N/2$ times, so that there are $N$ qubits between the operational qubits.
Each run of the algorithm always ends with the measurement of qubit $A$ and this information is saved on classical register $c$. To perform the second cycle of Deutsch's algorithm, the Pauli-X gate is applied on qubit $A$ only if 
the measurement outcome in the first cycle is $|1\rangle$.
Fig.~\ref{QcompDA2} shows an exemplary circuit  of Deutsch's algorithm in a situation where qubits $A$ and $B$ are swapped a total of $6$ times, and therefore there are 6 qubits between the operational qubits.

\begin{widetext}
\begin{center}
    \begin{figure}[!ht]
	\centering
   \includegraphics[width=1.\linewidth]{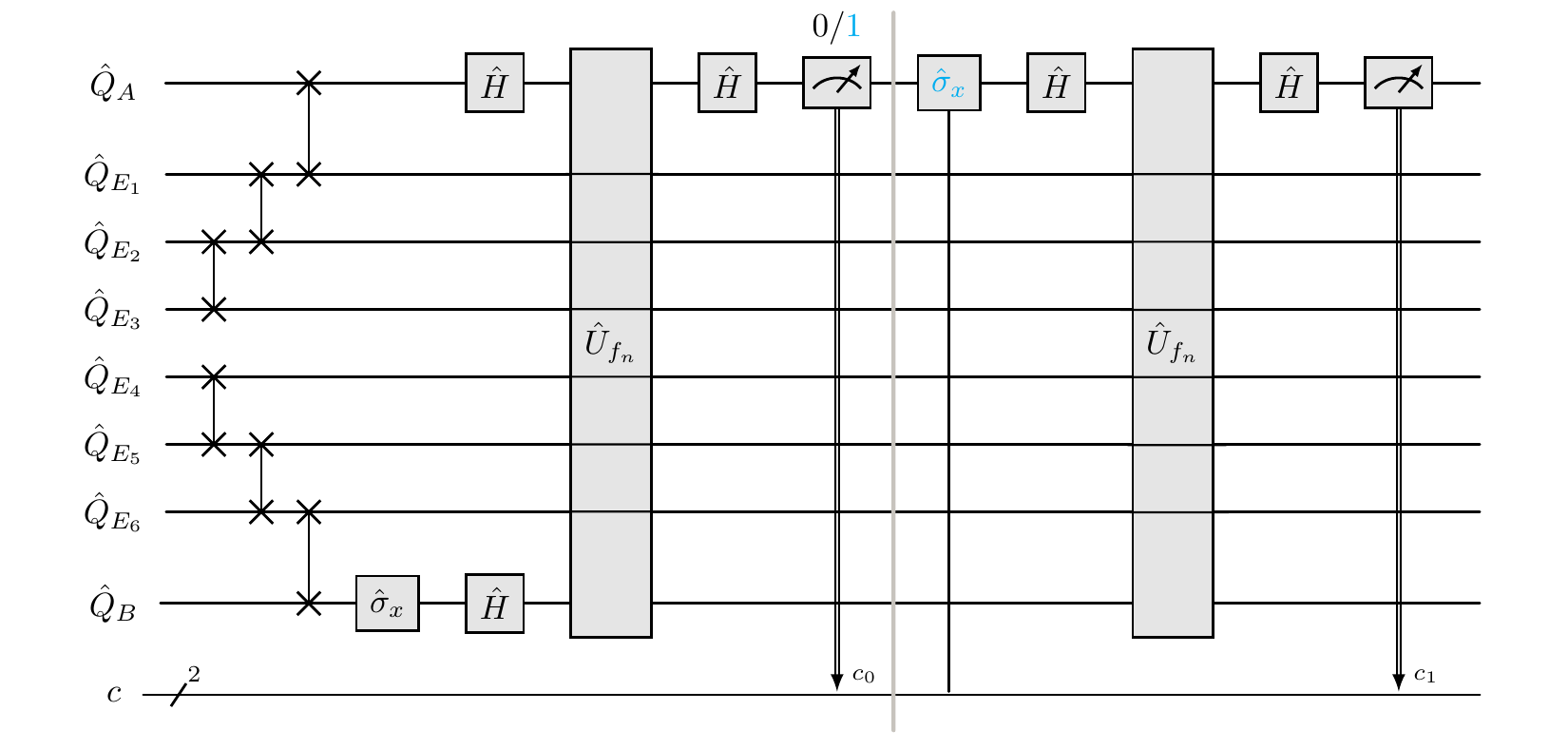}
	\caption{Flowchart of 2 cycles of Deutsch's algorithm for $N=6$ qubits beetween the operational qubits $A$ and $B$.
   Here swap gates are denoted by $x-x$, Pauli-X gates are marked by $\hat{\sigma}_x$, and Hadamard gates are symbolized by $\hat{H}$. The $U_{f_n}$ gates denote the two-qubit gate 
that only acts on the operational qubits. $c$ stands for the classical register. The vertical gray line denotes the end of the first cycle. 
		\label{QcompDA2}}
\end{figure}
\end{center}
\end{widetext}

\section{Details for NV center spin qubits}
Table \ref{table} contains the distances of the spinfull nuclei from the NV center spin qubit
and the corresponding values of the coupling constants used to calculate decoherence
for both qubits in Sec.~\ref{sec7}. 

\begin{figure}[!h]
    \centering
\includegraphics[width=1.\linewidth]{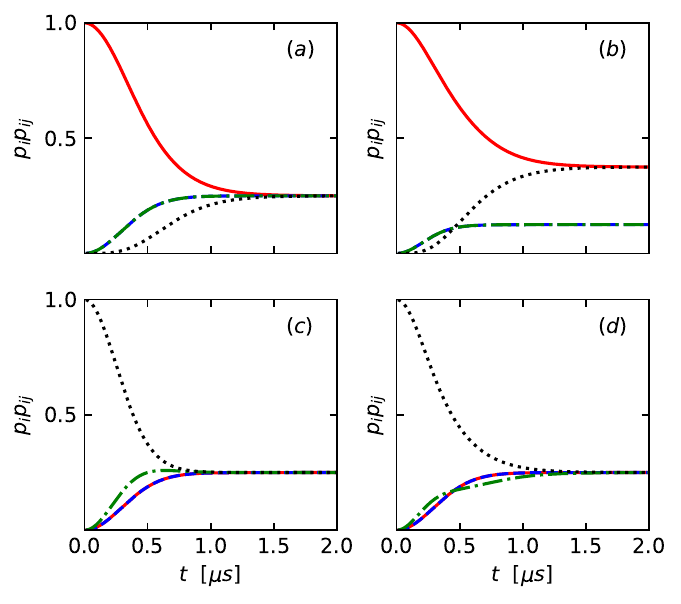}
    \caption{Probabilities of measurement outcome after two cycles of Deutsch's algorithm on NV center spin qubits as a function of decoherence time. 
    Initial polarization of nuclear environment is $p=0.1$ and applied magnetic field is $B_z=0.1$ T.
    Individual probabilities are given by: $p_{00}$ - solid red lines; $p_{01}$- dashed blue lines;  $p_{10}$ - dotted-dashed green lines; $p_{11}$ - dotted black lines. Left panels correspond to phase damping and right panels to quantum decoherence. Top panels show results for 
    constant functions and bottom for balanced.}
    \label{pjpji_p01b01}
\end{figure}

\begin{figure}[!h]
    \centering
\includegraphics[width=1.\linewidth]{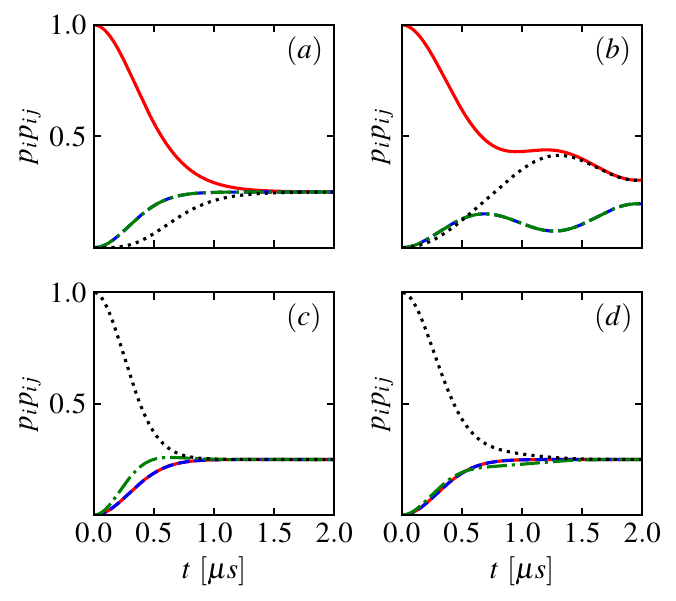}
    \caption{Same as in Fig \ref{pjpji_p01b01}, but for initially fully polarized nuclear spins, $p=1$}
    \label{pjpji_p1b01}
\end{figure}

Figs \ref{pjpji_p01b01} and \ref{pjpji_p1b01} are analogous to Figs \ref{p01b01} and \ref{p1b01}
in the main text and are made from the same data, but contain 
the dependence on decoherence time of the non-conditional probabilities.

\begin{table}
	\begin{tabular}{|c||c||c|c|c|}\hline
		$k$&$r_k$ [nm]&$\mathbb{A}^{z,x}_k$ [1/$\mu$s]&$\mathbb{A}^{z,y}_k$ [1/$\mu$s]&$\mathbb{A}^{z,z}_k$ [1/$\mu$s]\\
		\hline\hline
 $1$&$       0.527537	$&$-0.618725	$&$0.357221	$&$-0.631952$\\\hline
$2$&$ 0.563961$&$	0.196941$&$	-0.682223$&$	-0.417774$\\\hline
$3$&$0.636801$&$	-0.295048$&$-0.511038$&$	0.663824$\\\hline
$4$&$0.667287	$&$-0.382145$&$	0.220631$&$	0.660531$\\\hline
$5$&$0.667287$&$	0.339684$&$	0.147088$&$	-0.300241$\\\hline
$6$&$0.684928$&$	-0.372671$&$	-0.161371	$&$-0.22399$\\\hline
$7$&$0.756633$&$	0.0453058$&$	-0.39236$&$	-0.0320361$\\\hline
$8$&$0.756633$&$	-0.407753$&$	0	$&$0.288325$\\\hline
$9$&$0.756633$&$	-0.407753$&$	0	$&$0.288325$\\\hline
$10$&$0.756633$&$	0.407753$&$	0	$&$0.288325$\\\hline
$11$&$0.756633$&$	0$&$	0$&$	-0.288325$\\\hline
$12$&$0.756633$&$	-0.226529$&$	0$&$	-0.224253$\\\hline
$13$&$0.772235$&$	-0.138072$&$	0.079716$&$	-0.238655$\\\hline
$14$&$0.797561$&$	0.313331$&$	0.180902$&$	0.196941$\\\hline
$15$&$0.836489$&$	-0.0960167$&$	-0.11879$&$	-0.174585$\\\hline
$16$&$0.850628$&$	0.0819909$&$	0.284025$&$	0.173929$\\\hline
$17$&$0.873684$&$	0$&$	0$&$	-0.187272$\\\hline
$18$&$0.909359$&$	0.180677$&$	0.0625882	$&$0.24274$\\\hline
$19$&$0.909359$&$0.0993722$&$	-0.0469412	$&$-0.140533$\\\hline
$20$&$0.909359$&$	0.180677$&$	-0.0625882	$&$0.24274$\\\hline
$21$&$0.909359$&$	-0.0361353$&$	-0.187765$&$	0.24274$\\\hline
$22$&$0.909359$&$	-0.0813045$&$	0.234706$&$	0.0638789$\\\hline
$23$&$0.909359$&$	-0.0813045$&$	-0.234706	$&$0.0638789$\\\hline
$24$&$0.909359$&$	0.0722707$&$	0.187765$&$	-0.0638789$\\\hline
$25$&$0.922382$&$	-0.115691$&$	-0.200382$&$	0.0208233$\\\hline
$26$&$0.956242$&$	-0.154573$&$	0.133864	$&$0.00745225$\\\hline
$27$&$0.956242$&$	0.0298607$&$	-0.155161$&$	0.216115$\\\hline
$28$&$0.976808$&$	0.126338$&$	-0.0875292	$&$-0.0625339$\\\hline
$29$&$0.988943$&$	0.0801727$&$	0.138863$&$	-0.0440927$\\\hline
$30$&$0.988943$&$	-0.0282089$&$	-0.0488593	$&$0.249859$\\\hline
$31$&$0.988943$&$	0.0965042$&$	-0.0128577	$&$-0.102883$\\\hline
$32$&$0.988943$&$	-0.0593872$&$	0.0771462$&$	-0.102883$\\\hline
	\end{tabular}
	\caption{Table of distances of 32 environmental spin from the NV-center qubit
		and the corresponding coupling constants for the realization of the environment used in these paper.\label{table}}
\end{table}

\end{document}